\documentclass[reprint, onecolumn, aps, amsmath, amssymb,prd]{revtex4-2}
\usepackage{verbatim}
\usepackage{fullpage}
\usepackage{graphicx}
\usepackage{natbib}

\usepackage{bm}

\usepackage[dvipsnames]{xcolor}
\usepackage[colorlinks = true,
            linkcolor=BlueViolet,
            citecolor=OliveGreen,
            urlcolor=RedViolet
]{hyperref}
\usepackage[capitalise]{cleveref}

\usepackage{subcaption}

\usepackage{soul}

\newcommand{\refeq}[1]{Eq.~(\ref{eq:#1})}

\newcommand{\refapp}[1]{App.~\ref{app:#1}}

\def\apjl{Astrophys.\ J.\ Lett.}
\def\mnras{Mon.\ Not.\ R.\ Astron.\ Soc.}

\def\prd{Phys.\ Rev.\ D}

\def\physrep{Phys. Rep.}

\newcommand{\be}{\begin{equation}}
\newcommand{\ee}{\end{equation}}
\newcommand{\bea}{\begin{eqnarray}}
\newcommand{\eea}{\end{eqnarray}}
\newcommand{\vs}{\nonumber\\} 
\def\ba#1\ea{\begin{align}#1\end{align}}

\newcommand{\<}{\langle}
\renewcommand{\>}{\rangle}
\renewcommand{\[}{\left[}
\renewcommand{\]}{\right]}
\renewcommand{\(}{\left(}
\renewcommand{\)}{\right)}

\renewcommand{\ln}{\operatorname{ln}}

\renewcommand{\v}[1]{\bm{#1}}

\newcommand{\vk}{\bm{k}}

\newcommand{\vq}{\bm{q}}

\newcommand{\vx}{\bm{x}}

\newcommand{\bfx}{\bm{x}}

\newcommand{\bfk}{\bm{k}}

\newcommand{\vkhat}{\widehat{\v{k}}}

\newcommand{\D}{\Delta}

\renewcommand{\d}{\delta}

\newcommand{\eps}{\varepsilon}

\def\tt{\tilde t_{0}}

\def\dij{\delta_{ij}}
\def\hij{{h_{ij}}}

\def\D{\Delta}
\def\Dh2{\D_{\rm h}^2}

\def\ba#1\ea{\begin{align}#1\end{align}}

\def\({\left(}
\def\){\right)}

\def\[{\left[}
\def\]{\right]}

\def\vx{\bm{x}}
\def\vq{\bm{q}}
\def\bfx{\bm{x}}

\def\vk{{\bm{k}}}

\def\c{{\rm c}}
\def\b{{\rm b}}
\def\bc{\rm {bc}}

\def\khat{\widehat{\bm{k}}}

\def\qhat{\widehat{\bm{q}}}

\def\cG{{\cal G}}
\def\gf{{\cal G}(k, \eta, \widetilde \eta)}

\begin{document}

\title{Gauge-Invariant Tensor Perturbations Induced from Baryon-CDM Relative Velocity and the B-mode Polarization of the CMB}

\author{James Gurian }
\affiliation{Department of Astronomy and Astrophysics and Institute for Gravitation and the Cosmos, \\ The Pennsylvania State University, University Park, PA 16802, USA}

\author{Donghui Jeong}
\affiliation{Department of Astronomy and Astrophysics and Institute for Gravitation and the Cosmos, \\ The Pennsylvania State University, University Park, PA 16802, USA}

\author{Jai-chan Hwang}
\affiliation{Department of Astronomy and Atmospheric Sciences, Kyungpook National University, Daegu, 702-701, Korea}
\affiliation{Center for Theoretical Physics of the Universe, Institute for Basic Science (IBS), Daejeon, 34051, Korea}

\author{Hyerim Noh}
\affiliation{Korea Astronomy and Space Science Institute, Daejeon, 305- 348, Korea}

\begin{abstract}
At second order, scalar perturbations can source traceless and transverse perturbations to the metric, called induced gravitational waves (IGW). The apparent gauge-dependence of the IGW obscures the interpretation of the stochastic gravitational-wave signal. To elucidate the gauge dependence, we study the IGW from manifestly gauge-invariant scalar perturbations, namely, the relative velocity between baryons and cold dark matter. From this relative velocity perturbation, we compute the dimensionless gravitational wave power spectrum and the corresponding expected angular power spectrum of the B-mode polarization of the cosmic microwave background. Although the effect turns out to be unobservably small, the calculation demonstrates both the importance of using observable quantities to remove the gauge ambiguity and the observable consequences of tensor perturbations which are not propagating gravitational waves.
\end{abstract}
\maketitle
\section{Introduction}

In general relativity, perturbations of the space-time metric can be decomposed into scalar, vector, and tensor (SVT) types according to their transformation properties under spatial coordinate transformation \cite{bardeengauge-invariant1980}. 
At linear order, these three types of perturbations evolve independently, but at second order the nonlinearity of the Einstein equations couples the SVT perturbations to each other \cite{Noh2004}. This means that for example an initially scalar perturbation does not remain purely scalar as it evolves, and the decomposition depends on the choice of a constant-time hypersurface slicing (a gauge choice). 

In particular, scalar perturbations can couple at second order to generate traceless and transverse tensor perturbations, known as induced gravitational waves (IGW) \cite{mollerachcmb2004, Baumann2007}. These IGW effects are important because they will contribute to the stochastic gravitational wave (SGW) background. The SGW background is generated by sources including close binary systems and supermassive black hole binary mergers \cite{Christensen2018}. The large-scale (low-frequency) components of the SGW signal can also be generated by primordial black holes \cite{Bugaev2010, Chen2020}. A confirmed detection of SGW background would open up a new window to these ubiquitous GW sources. LIGO has placed an upper bound \cite{Abbott2017} on kilohertz frequency SGW. In the nanohertz regime, pulsar timing arrays (PTA) can measure the SGW background \cite{alam2020}. Recently, the NANOgrav pulsar timing project detected a stochastic signal in their 12.5 year data, but due to the lack of quadrupolar correlations they do not claim this as a detection of the SGW background \cite{arzoumanian2020nanograv}.

However, the gauge dependence of the IGW complicates any attempts to physically interpret a future definitive SGW detection. The issue of gauge freedom in cosmological perturbation theory is well described in the literature \cite{bardeengauge-invariant1980, bardeen1988, mukhanovtheory1992, pitrouradiative2009, nakamurasecond-order2007}. In brief, the gauge freedom in general relativity corresponds to smooth changes of coordinates. All observables must be preserved under gauge transformations, but a perturbation variable is not itself an observable. The reason is that defining a perturbation requires defining a background and decomposing all quantities into their background and perturbed values. The full quantity is physical, but the decomposition is arbitrary. This means that the value of the perturbation depends on the choice of background (gauge). Since the decomposition into perturbations and background is not in general preserved under changes of coordinates, perturbation variables can depend on gauge. 

It is possible to define gauge-invariant variables from the perturbation variables in a particular gauge, but this does not resolve the issue. The reason is that the interpretation of the variable depends on the gauge. For example the Bardeen potential $\Psi$ is gauge invariant, but can be interpreted as a curvature perturbation only when the timelike hypersurfaces are chosen such that there is no shear between normal worldlines \cite{bardeen1988}. Said another way, defining a gauge independent variable from a gauge dependent variable is a process of coming to an agreement about how a particular observable will appear in a particular system of coordinates. In order to calculate a quantity as it would be measured, we must begin with the frame of reference relevant to the source and end with the frame relevant to the detector. 

In the case of the IGW sourced from the linear-order density perturbation, the gauge issue has been extensively discussed but is not yet settled. The authors of Ref.~\cite{hwanggauge2017} calculated wildly different gravitational wave energy densities in different gauges for a source in the matter-dominated epoch. Due to its relevance to the primordial black hole problem, most follow-up studies have focused on sources in the radiation-dominated epoch. The authors of
Refs.~\cite{tomikawagauge2019, inomatagauge2020, Yuan2020, Lu2020} have repeated this calculation in several gauges and (unsurprisingly) have found agreement between some gauges but not others. Most comprehensively, Ref.~\cite{gong2019analytic} derived analytic solutions for the induced gravitational waves from all the second-order source terms in multiple gauges during both the matter and the radiation dominated epochs. Meanwhile, Ref.~\cite{Luca2020} attempts to carry out the calculation in a
gauge-invariant way. That work argues that the traceless-transverse gauge is relevant to detection by LISA and that the gravitational wave energy density in this gauge coincides with the energy density calculated from the gauge-invariant variables defined by the Poisson gauge. However the argument is based on a calculation only at first order in the perturbations. Moreover, the coincidence between the Poisson gauge result and traceless-transverse gauge result relies on the rapid decay of the
gravitational potential inside the horizon during the radiation dominated epoch. However, this mathematical observation alone cannot replace physical justification for a particular gauge choice.Ref.~\cite{domenech2020approximate} claimed that the IGW in the dust era could be ``gauged away''. However, the fact that the gauge dependent IGW can be eliminated by a gauge transformation does not necessarily imply that the IGW is physically irrelevant. As we show in the
\refapp{app}, as long
as the IGW is gauge dependent to second-order, we can write down a gauge condition (on the linear scalar perturbation) that cancels the IGW completely. We may call it the ``zero-IGW gauge.''  
As we argue in the \refapp{app}, the existence of zero-IGW gauge is not related to the status of the IGW as an observable. The correct gauge should be determined based on the observational strategy, as Refs.~\cite{Yoo2009,Challinor2011, Jeong2012} did for the matter power spectrum. Those works showed that in fact none of the previously known gauge conditions are relevant for the observed galaxy power spectrum. Likewise, it will be necessary to construct the correct gauge choice for the gauge-dependent IGW based on the nature of the production, propagation, and detection method.

\begin{figure}
{
    \begin{subfigure}{.5\textwidth}
    \includegraphics[width=\linewidth]{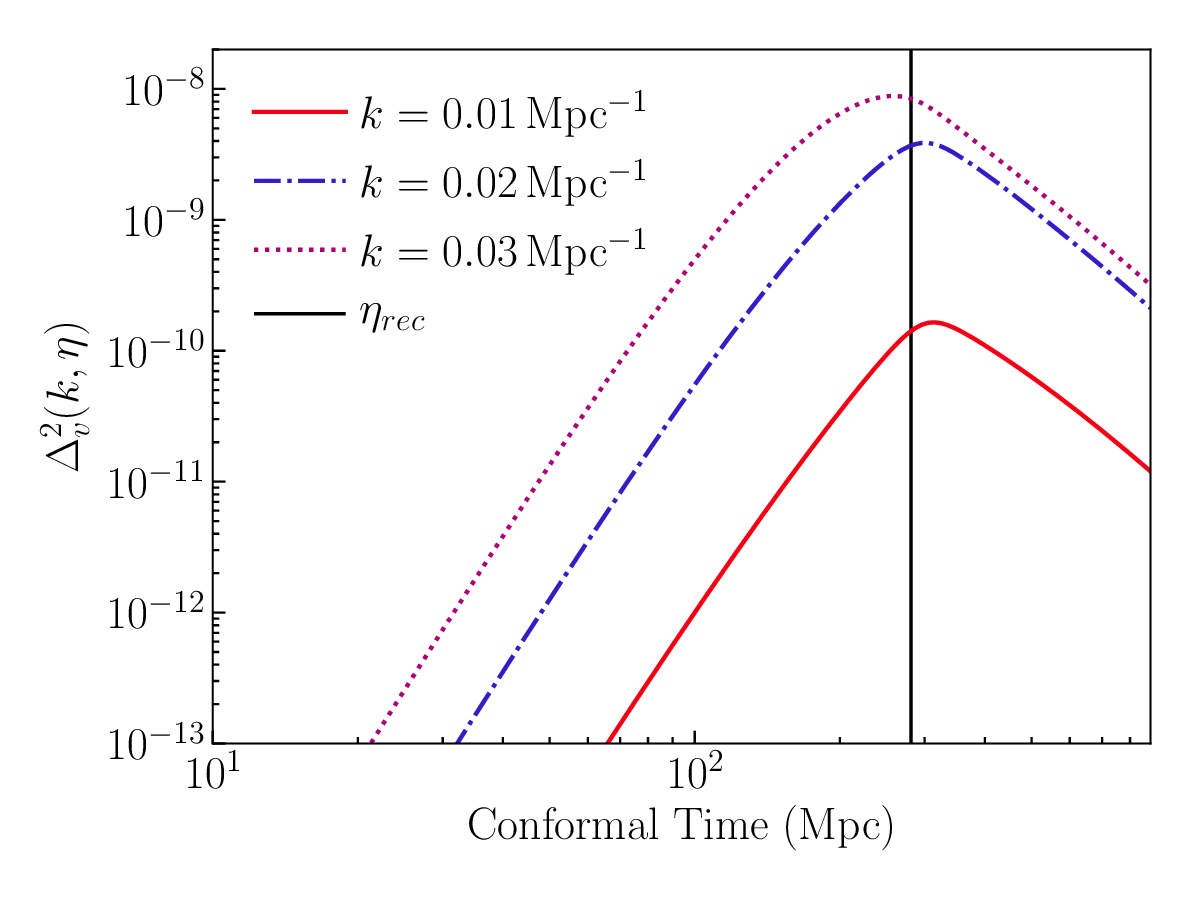}
    \end{subfigure}%
    \begin{subfigure}{.5\textwidth}
    \includegraphics[width=\linewidth]{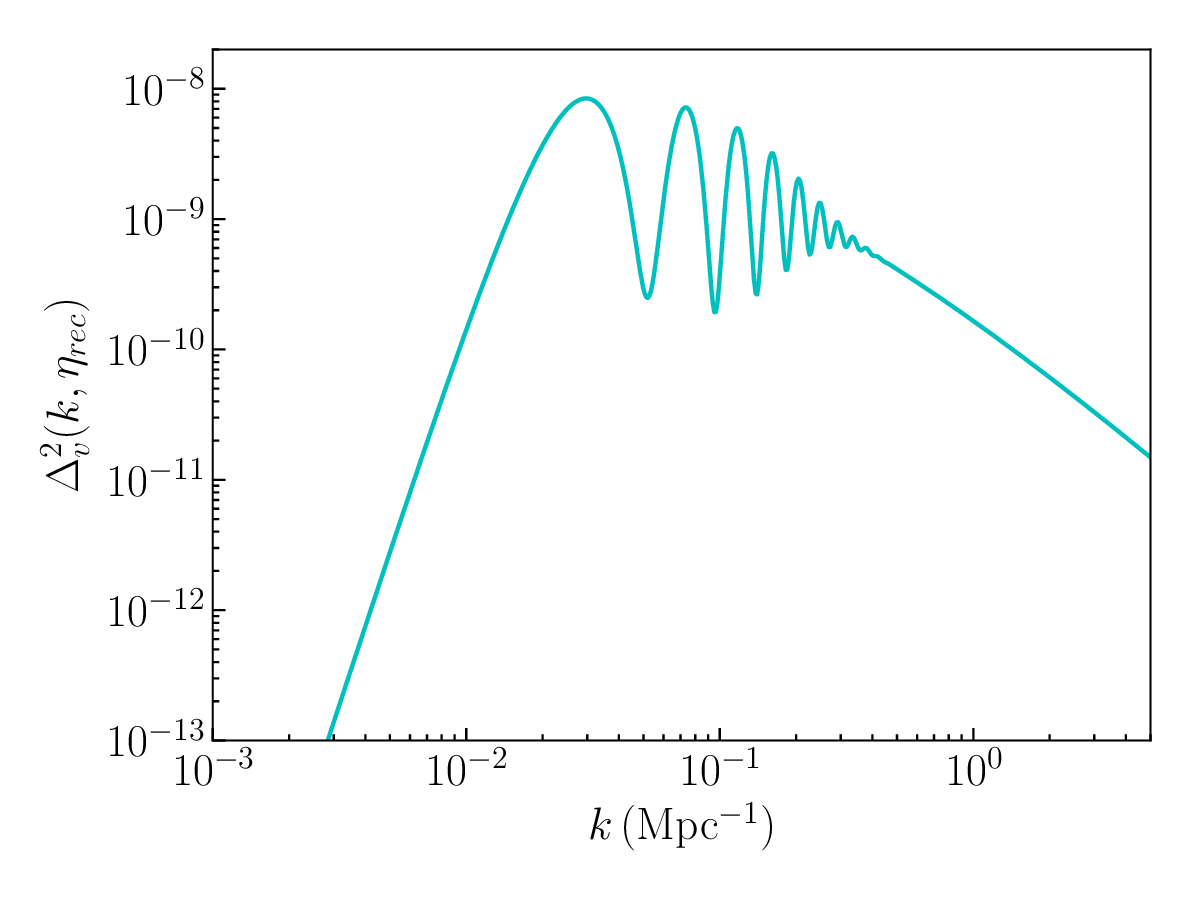}
    \end{subfigure}
}
\captionsetup{singlelinecheck = false,  justification=raggedright, labelsep=space}
\caption{\textbf{Left:} 
    The evolution of the relative velocity variance per $\ln k$ for three relevant modes in units of $c$, defined by $\Delta^2_v(k) = \frac{k^3}{2\pi^2} T_{bc}^2(k, \eta)P_\zeta(k)$, where $T_{bc}$ is the relative velocity transfer function and $P_\zeta$ is power spectrum of the primordial curvature perturbation. The vertical line shows $\eta_{*}$ of CMB last-scattering surface. 
    \textbf{Right:} The relative velocity variance per $\ln k$ at recombination. 
    \label{fig:rv}
}
\end{figure}

Here, we calculate the IGW sourced not from density perturbations but instead from relative velocity perturbations. A relative velocity IGW effect was also mentioned (and shown to be negligible at the scales of interest to that work) in Ref.~\cite{Dom_nech_2021}, there between radiation and primordial black hole fluids. In this work we consider the relative velocity between the photon-baryon fluid and the CDM (cold dark matter) prior to the cosmic recombination. The existence of this effect is long recognized \cite{BondCMB}, and has been previously studied in the context of the formation of the first structures \cite{tseliakhovichrelative2010}. The idea is that the baryon-photon perturbations are supported against gravitational collapse by pressure, which leads to the formation of standing acoustic waves. Meanwhile, the standard {\it cold} dark matter perturbation grows independent of the  baryon-photon pressure. Once a perturbation enters the horizon a relative velocity builds up between the two components, and then decays proportional to the reciprocal of the scale factor. The relative velocity is dominated by the scalar (longitudinal) part, and root-mean-squared amplitude of the relative velocity is $\sqrt{\<(v/c)^2\>} \sim 10^{-4}$ at recombination time. We show the time evolution of the relative velocity variance per $\ln k$ for three relevant modes ({\it left panel}) and the relative velocity variance per $\ln k$ at recombination over a wide range of scales ({\it right panel}) in  FIG.~\ref{fig:rv}. 

At second order, the relative velocity couples to itself, producing a traceless-transverse (anisotropic stress) component in the energy momentum tensor that sources IGW \cite{Hwang-Noh-Park-2016}. In this work, we investigate the properties of relative-velocity IGW through its contribution first to the tensor power spectrum $P_h(k)$ and then to the B-mode power spectrum of the cosmic microwave background (CMB) polarization.

The calculation of the relative velocity IGW parallels that of the density perturbation IGW effect, except that in this case we begin with the physical relative velocity rather than an arbitrary perturbation variable. Therefore, the result is naturally gauge invariant. The gauge-invariance of the source makes the relative velocity IGW a comparatively straightforward example of a calculation in higher order relativistic perturbation theory which starts and ends with gauge invariant quantities. We begin with the physically unambiguous baryon-CDM relative velocity and end with the directly observable polarization on the sky. Quantities which could in principle depend on gauge appear only in the middle part of the calculation. This calculation also clarifies that the vacuum ``freely propagating'' gravitational wave solutions do not have a privileged position as observables. Even while the source term is still active and the solution does not at all resemble freely propagating gravitational radiation, we show that anisotropic stress still has physical consequences.

The paper is organized as follows. In section II, we show the contribution of the relative velocity to the stress-energy tensor at second order, and then calculate the power spectrum of the relative velocity IGW, $P_h(k)$ using the method of Green's functions. In section III we calculate the B-mode power spectrum arising from this effect in the analytic approximation of Ref.~\cite{pritchardcosmic2005}. We conclude with a summary of our results and discussion for the future directions of the IGW study.

We use the Planck 2015 best-fitting cosmological parameters for flat-$\Lambda$CDM cosmology \cite{Planck}: $\Omega_b h^2 = 0.0223$, $\Omega_m = 0.309$, $\Omega_\Lambda = 0.691$, $H_0 = 67.74$ km/s/Mpc.

\section{Calculation of the Relative Velocity Induced Gravitational Waves}
\subsection{The second-order energy momentum tensor}
For multi-component fluid, the collective energy-momentum tensor and the fluid quantities in terms of individual ones valid to fully nonlinear order perturbation are presented in Section 3.2 of Ref.~\cite{Hwang-Noh-Park-2016}. We only need the anisotropic stress that sources IGW. To the second order, from (33) and (35) of Ref.~\cite{Hwang-Noh-Park-2016} we have
\bea
   & & \Pi_{ij} 
       = \sum_K \left( \rho_K + {1 \over c^2} p_K \right)
       \left[ \left( v_i - v_{Ki} \right)
       \left( v_j - v_{Kj} \right)
       - {1 \over 3} \delta_{ij}
       \left( v^k - v_{K}^k \right)
       \left( v_k - v_{Kk} \right) \right],
   \nonumber \\
   & & v_i = {\sum_K \left( \rho_K + p_K/c^2 \right) v_{Ki}
       \over \sum_J \left( \rho_J + p_J/c^2 \right)}, 
       \label{eq:vi}
\eea
where the subscript $K$ runs over the components. Note that we have ignored the anisotropic stress of individual component. The above anisotropic stress is generated by velocity differences among fluid components. The anisotropic stress tensor $\Pi_{ij}$ and the velocity vector $v_i$ are introduced, to the second-order perturbation, as
\bea
   & &   \Pi_{ij} \equiv \frac{\widetilde\pi_{ij}}{a^2}  , \quad
        {v_i } \equiv \frac{\widetilde u_ic}{a} , \quad
        {v_{Ii} } \equiv  \frac{\widetilde u_{Ii}c}{a} ,
\eea
where $\widetilde \pi_{ab}$ and $\widetilde u_a$ are the covariant anisotropic stress and the fluid four-vector for collective fluid, and $\widetilde u_{Ia}$ is the fluid four-vector for individual component under the energy-frame condition, $\widetilde q_{Ia} \equiv 0$; indices $a, b \dots$ and $i, j \dots$ indicate space-time and space, respectively. The indices of $\Pi_{ij}$ and $v_i$ are raised and lowered using, in the spatially flat background, $\delta_{ij}$ as the metric, and $v_i$ in \refeq{vi} follows from the energy frame condition $\widetilde q_a \equiv 0$.

For a two fluid system with baryon ($\b$) and CDM ($\c$) we have (setting from here on $c \equiv 1$)
\bea
   & & \Pi_{ij} = {\rho_\c \left( \rho_\b + p_\b \right)
       \over \rho_\c + \rho_\b + p_\b}
       \left[ \left( v_{\c i} - v_{\b i} \right)
       \left( v_{\c j} - v_{\b j} \right)
       - {1 \over 3} \delta_{ij}
       \left( v_{\c}^k - v_{\b}^k \right)
       \left( v_{\c k} - v_{\b k} \right) \right].
   \label{Pi_ij}
\eea
To the second order, we need $v_{\bc i} \equiv v_{\c i} - v_{\b i}$ only to the linear order. To the linear order, under gauge transformation $\widehat x^a = x^a + \xi^a (x^e)$, we have $\widehat v_{Ii} = v_{Ii} + \xi^0_{,i}$, thus $v_{\c i} - v_{\b i}$ is gauge invariant. This proves that $\Pi_{ij}$ is gauge invariant.

A transverse-tracefree (TT) projection of $\Pi_{ij}$ contributes to the IGW. From (8) of Ref.~\cite{hwanggauge2017}, the projection operator is given as
\ba
     \Pi_{ij}^{\rm TT} &\equiv {\cal P}_{ij}^{\;\;\;k\ell}\Pi_{k\ell}\\
       & \equiv \Pi_{ij}
       + {1 \over 2} \left( \Delta^{-1} \nabla_i \nabla_j
       - \delta_{ij} \right) \Pi^k_k
       - \Delta^{-1} \left( \nabla_i \Pi^k_{j,k}
       + \nabla_j \Pi^k_{i,k} \right)
       + {1 \over 2} \Delta^{-1} 
       \left( \Delta^{-1} \nabla_i \nabla_j
       + \delta_{ij} \right) \Pi^{k\ell}_{\;\;\;,k\ell}\,.
       \label{TT}
\ea

\subsection{The induced gravitational waves}
Let us turn our attention to the IGW from the anisotropic stress term that we calculated in the previous section. Throughout, we define
\be
\int_{\vk}\equiv \int\frac{d^3k}{(2\pi)^3}.
\ee
We focus on the traceless-transverse part $h_{ij}$ of the metric perturbation:
\be
ds^2 = a^2(\eta)\[-d\eta^2 + \(\dij + \hij(\bfx, \eta)\)dx^idx^j \]\,.
\label{eq:metric}
\ee

The equation of motion for the tensor perturbations is \cite{Lifshitz46, bardeengauge-invariant1980}
\be
h''_{ij}(\vx, \eta) + 2{\cal H} h'_{ij}(\vx, \eta) - \nabla^2 h_{ij}(\vx, \eta) = 16 \pi G a^2 \Pi_{ij}^{\rm TT}(\vx, \eta),
\ee
where prime denotes derivatives with respect to conformal time, and ${\cal H}\equiv a'/a=aH$.

In the spatially flat Universe, it is most natural to study the perturbation in the Fourier basis defined by 
\be
\hij(\vx, \eta)
=
\sum_{p} \int_{\vk} h_p(\vk, \eta) \, 
{\eps^p}_{ij}(\vkhat) \,
e^{i\vk\cdot\vx},
\label{eq:Fourier}
\ee
where the subscript $p\in\{+,\times\}$ indicates one of the polarization bases for two tensor (transverse-traceless) modes. We normalize the polarization basis as: 
\be
{\eps^p}_{ij}\eps^{p',ij} = 2 \d^{pp'}.
\ee
Then, the Einstein equation becomes the sourced ordinary differential equation for the Fourier components, 
\be
h_p''(\vk, \eta) + 2{\cal H}h_p'(\vk, \eta) + k^2 h_p(\vk, \eta)= s_p(\vk, \eta)\,,
\label{eq:einseq}
\ee 
with the source term given as 
\begin{equation}
    s_p(\vk, \eta) = 16\pi G a^2(\eta) \int_{\bfx} \Pi_{ij}^{TT}(\vx, \eta)\, \eps^{p, ij}(\vkhat) \,
e^{-i\bfk\cdot\bfx}.
\end{equation}

For the IGW sourced by baryon-CDM relative velocity, we can derive the following expression for the source term by using (\ref{Pi_ij}) and (\ref{TT}):
\ba
s_p(\vk,\eta) =
\rho_{\bc}(\eta)a^2(\eta)
\int_{\vq} {\eps}^{p,ij}(\widehat{\vk}) (\widehat{\vk-\vq})_i \widehat\vq_j T_{\bc}(q,\eta)T_{\bc}(|\vk -\vq|,\eta) \zeta(\vq) \zeta(\vk-\vq)\,.
\label{eq:sp}
\ea
Here, for notational convenience we have defined the quantity $\rho_{bc}$ by 
\be
\rho_{\bc}(\eta) =  16 \pi G \frac{\rho_\c(\eta)\left(\rho_\b(\eta) + p_\b(\eta)\right) }{\rho_\c(\eta) +\rho_\b(\eta) + p_\b(\eta)}\,,
\ee 
and we have introduced the transfer function $T_{\bc}(k)$ for the relative velocity ${\bm v}_{\bc}$ through the relation 
\be 
{\bm v}_{\bc}(\vk, \eta) = \widehat{\vk}T_{\bc}(\vk, \eta)\zeta(\vk),
\ee
with $\zeta(\vk)$ being the primordial curvature perturbation. 

Now, we recast \eqref{eq:einseq} in terms of $v_p(\vk,\eta)= a(\eta)h_p(\vk,\eta)$ to simplify the derivation of the analytic solutions in the next two sections. The equation for $v_p(\vk, \eta)$ is

\begin{equation}
    v_p''(\vk, \eta) + \left[k^2-\frac{a''(\eta)}{a(\eta)}\right]v_p(\vk, \eta) = a(\eta)s_p(\vk, \eta).\label{ve}
\end{equation}

We solve this using the method of Green's functions, where the general solution to \eqref{ve} is given in terms of the Green's function ${\cal G}$ and homogeneous solutions $v_1$ and $v_2$ as
\be
v_p(\vk, \eta) = c_{1p}(\vk) v_1(k, \eta) + c_{2p}(\vk) v_2(k, \eta) + \int_0^\eta d\widetilde\eta\, a(\widetilde\eta)s_p(\vk, \widetilde\eta){\cal G}(k, \eta, \widetilde \eta).
\label{eq:vsol}
\ee

\subsection{Homogeneous solutions in single-component universes}
We first present the solutions to \eqref{ve} for the single-component matter and radiation dominated universes in order to clarify our approach and develop intuition concerning the full solution in our Universe.

In the radiation dominated epoch $a(\eta) \propto \eta$, so that 
the homogeneous part of \eqref{ve} reduces to 
\be
v''_p(k, \eta) + k^2 v_p(k, \eta) = 0\,,
\ee
and the two homogeneous solutions are
\begin{gather}
    v_1(k, \eta) = k \eta j_0(k\eta)\\
    v_2(k, \eta) = k \eta y_0(k\eta).
\end{gather}
Meanwhile, during the matter dominated epoch $a(\eta)\propto \eta^{2}$. Then, the homogeneous part of \eqref{ve} becomes
\be
\eta^2 v''_p(k\eta) + [(k\eta)^2 -2 ]v_p(k, \eta)=0,
\ee
with homogeneous solutions
\begin{gather}
    v_1(k, \eta) = k \eta j_1(k\eta)\\
    v_2(k, \eta) = k \eta y_1(k\eta).
\end{gather} 

In terms of these homogeneous solutions, the Green's function is given as
\be
\gf \equiv \frac{v_1(k, \eta) v_2(k, \widetilde\eta) - v_1(k, \widetilde\eta) v_2(k, \eta)}{v_1'(k, \widetilde\eta) v_2(k, \widetilde\eta) - v_1(k, \widetilde\eta)  v_2'(k, \widetilde\eta)}.
\label{gf}
\ee

\subsection{Numerical Solution for Two Component Universe}
The wave modes that generate the relative velocity IGW relevant for B-mode polarization power spectrum enter the comoving horizon well before the matter-radiation equality. Therefore, the IGW we consider here begins to grow before the epoch of matter-radiation equality, and we must consider the two component universe containing both matter and radiation. In this case, the time evolution of the scale factor $y = a/a_{eq}$ can be found from the Friedmann equation at matter-radiation equality as
\be
\frac{dy}{d\eta} = k_{eq}\sqrt{\frac{y+1}{2}},
\ee
which has solution 
\be
y = \frac{1}{\sqrt2}k_{\rm eq}\eta + \frac18 k_{\rm eq}^2 \eta^2.
\ee
Here, 
$k_{\rm eq}\equiv a_{\rm eq}H_{\rm eq} \simeq 0.073\,\Omega_{\rm m} h^2\,{\rm Mpc^{-1}} \simeq 0.0104 \,\rm Mpc^{-1}$ is the wavenumber of the comoving horizon at the matter-radiation equality.
Substituting this result into \eqref{ve} gives 
\be
v''_p + \left[k^2 -\frac{2\sqrt2 k_{\rm eq}}{\eta(\sqrt2 k_{\rm eq} \eta + 8)}\right]v_p = 0\,,
\ee
for which we obtain the homogeneous solutions numerically. We also calculate the Green's function $\gf$ using the numerical solutions in the following way. First, we recall that the Green's function is defined as the response of a system at advanced time $\eta$ to an impulse at retarded time $\widetilde{\eta}$. That is, the Green's function for a differential operator ${\cal L}_\eta$ solves 
\be
{{\cal L}_\eta  {\cal G}}(k, \eta, \widetilde\eta) = \delta^D(\eta - \widetilde\eta).
\label{eq:Greensfn}
\ee
So, to find ${\cal G}(k, \eta, \widetilde\eta)$  for fixed $\widetilde\eta$ as a function of $\eta$, we numerically solve the homogeneous equation on the interval $[\widetilde\eta, \eta_{\rm max}]$ subject to the initial conditions imposed by a delta function impulse at $\widetilde\eta$. That is, 
\begin{gather} 
{\cal G}(k, \eta = \widetilde\eta) = 0\\
{\cal G}'(k, \eta = \widetilde\eta) = 1.
\end{gather}
In the single component universe, applying this definition leads to the form of the Green's function \eqref{gf}.

For this calculation, however, we want to know instead how an impulse at time $\widetilde\eta$ propagates to some fixed $\eta_{\rm max}$; ${\cal G}(k, \eta_{\rm max}, \widetilde\eta)$ as a function of $\widetilde\eta$. Therefore, it is computationally preferable to impose the boundary conditions at $\widetilde\eta = \eta_{\rm max}$, finding the advanced Green's function ${\cal G} (k, \eta, \eta_{\rm max})$ and then use the anti-symmetry of the Green's functions 
\be
{\cal G}(k,\eta,\eta') = 
-
{\cal G}(k,\eta',\eta)\,,
\label{eq:anti-G}
\ee
(which is apparent in \eqref{gf}), to exchange the arguments. This gives us all needed values of the Green's function for a given $k$ by integrating only a single differential equation.
We have tested that the Green's function calculated in this way coincides with eq. (24) for $\eta$, $\eta'$ during the radiation epoch.

\subsection{IGW Power Spectrum}
We are now ready to compute the tensor power spectrum, defined by 
\be
4\langle h_p(\vk,\eta) h_p(\vk',\eta)\rangle =
(2\pi)^3 \delta^D(\vk+\vk')P_h(\vk,\eta)  
\ee
where the factor of 4 comes from statistical isotropy in the polarization (i.e.~a factor of 2 from each polarization, in our normalization). 

Using the Green's function ${\cal G}(k,\eta,\eta')$ for $v_p(\vk,\eta)=a(\eta)h_p(\vk,\eta)$ defined in \refeq{Greensfn} and the traceless-transverse source term $s_p({\bm k},\eta)$ in \refeq{sp}, we may write the power spectrum of the IGW as 
\begin{align}
&(2\pi)^3 \delta^D(\vk+\vk')P_h(\vk,\eta) 
\vs
=& \frac{4}{a^2(\eta)} \int_0^\eta d\eta'\,a(\eta'){\cal G}(k, \eta, \eta') \int_0^\eta d\eta''\, a(\eta''){\cal G}(k', \eta, \eta'')  \langle s_p(\vk, \eta')s_p(\vk', \eta'')\rangle
\vs
=& \frac{4}{a^2(\eta)} \int_0^\eta d\eta'\,a^3(\eta'){\cal G}(k, \eta, \eta')\rho_{\bc}(\eta')\int_0^\eta d\eta''\,a^3(\eta''){\cal G}(k', \eta, \eta'') \rho_{\bc}(\eta'') 
\vs
&\quad\times
\int_{\vq_1}\int_{\vq_2}
{\eps}^{p,ij}(\widehat\vk)
\qhat_{1,i}
\widehat{(\vk-\vq_1)}_j
\eps^{p,lm}(\widehat\vk')
\qhat_{2,l}
\widehat{(\vk'-\vq_2)}_m
T_{\bc}(q_1,\eta')
T_{\bc}(|\vk -\vq_1|,\eta')
T_{\bc}(q_2,\eta'')T_{bc}(|\vk' -\vq_2|,\eta'')  
\vs
&\qquad\qquad\times
\langle
\zeta(\vq_1) \zeta(\vk-\vq_1)\zeta(\vq_2) \zeta(\vk'-\vq_2)
\rangle\,. 
\end{align}
We then evaluate the four-point correlator using Wick's theorem:
\be
\langle
\zeta(\vq_1) \zeta(\vk-\vq_1)\zeta(\vq_2) \zeta(\vk'-\vq_2)
\rangle
=
\langle
\zeta(\vq_1) 
\zeta(\vq_2) 
\rangle
\langle
\zeta(\vk-\vq_1)
\zeta(\vk'-\vq_2)
\rangle
+
\langle
\zeta(\vq_1) 
\zeta(\vk'-\vq_2)
\rangle
\langle
\zeta(\vq_2) 
\zeta(\vk-\vq_1)
\rangle\,,
\ee
but ignore the contribution which is only non-zero at $\vk=\vk'=0$.
Thanks to statistical isotropy, $P_h(\vk,\eta)=P_h(k,\eta)$, we are free to chose $\widehat{\vk} = \widehat{\bm{z}}$, and the expression reduces to 
\ba
P_h(k, \eta)  
&= \frac{8}{a^2(\eta)} \int_0^\eta d\eta'\,a^3(\eta'){\cal G}(k, \eta, \eta')\rho_{\bc}(\eta')\int_0^\eta d\eta''\,a^3(\eta''){\cal G}(k, \eta, \eta'') \rho_{\bc}(\eta'') 
\vs
&\quad\times
\int_{\vq}
\frac{\[{\eps}^{p,ij}(\widehat{\bm{z}})q_iq_j\]^2}{q^2|\vk-\vq|^2}
T_{\bc}(q,\eta')
T_{\bc}(|\vk -\vq|,\eta')
T_{\bc}(q,\eta'')T_{\bc}(|\vk -\vq|,\eta'')  
P_\zeta(q) P_\zeta(|\vk-\vq|)
\,.
\ea
Here, we use the transverse property of the tensor polarization to simplify
\be
{\eps}^{p,ij}(\widehat\vk)
\qhat_{i}
\widehat{(\vk-\vq)}_j
=
\frac{{\eps}^{p,ij}(\widehat\vk)q_{i}(k-q)_j}{q|\vk-\vq|}
=
-\frac{{\eps}^{p,ij}(\widehat\vk)q_{i}q_j}{q|\vk-\vq|}\,.
\ee
For the calculation, we choose $p=+$ polarization with which
\be
{\eps}^{p,ij}(\widehat{\bm{z}})q_iq_j
=
q_x^2 - q_y^2 
=
q^2 \sin^2\theta \cos(2\varphi)\,,
\ee
with, respectively, the polar and azimuthal angles $\theta$ and $\varphi$ in a coordinate where $\vk\parallel\widehat{\bm{z}}$. We evaluate the transfer functions $T_{\bc}(k,\eta)$ using CAMB \cite{Lewis2000}. The integration is performed using the Monte-Carlo routine Divonne, provided as part of the CUBA \cite{Hahn2005} library.
\begin{figure}
{   
    \begin{subfigure}{.5\textwidth}
    \includegraphics[width=\linewidth]{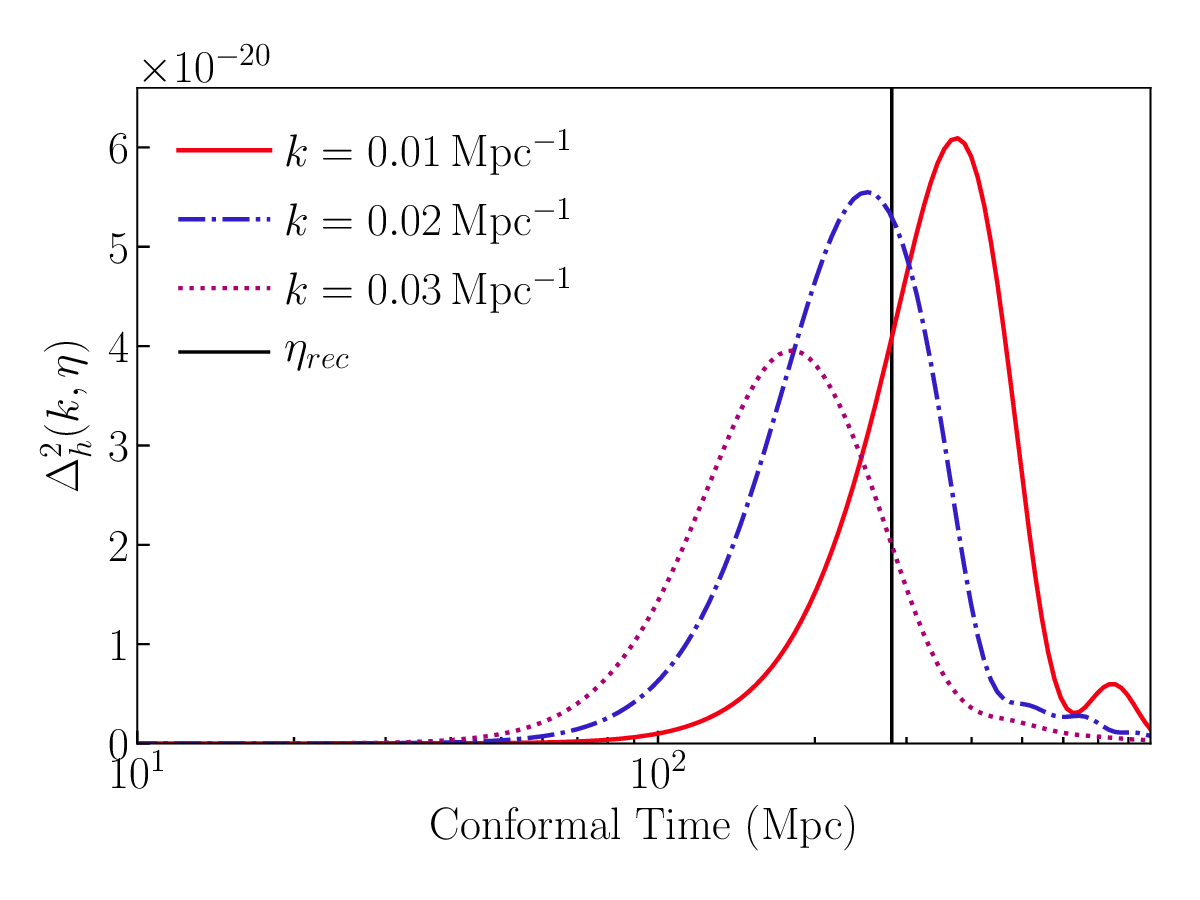}
    \end{subfigure}%
    \begin{subfigure}{.5\textwidth}
    \includegraphics[width=\linewidth]{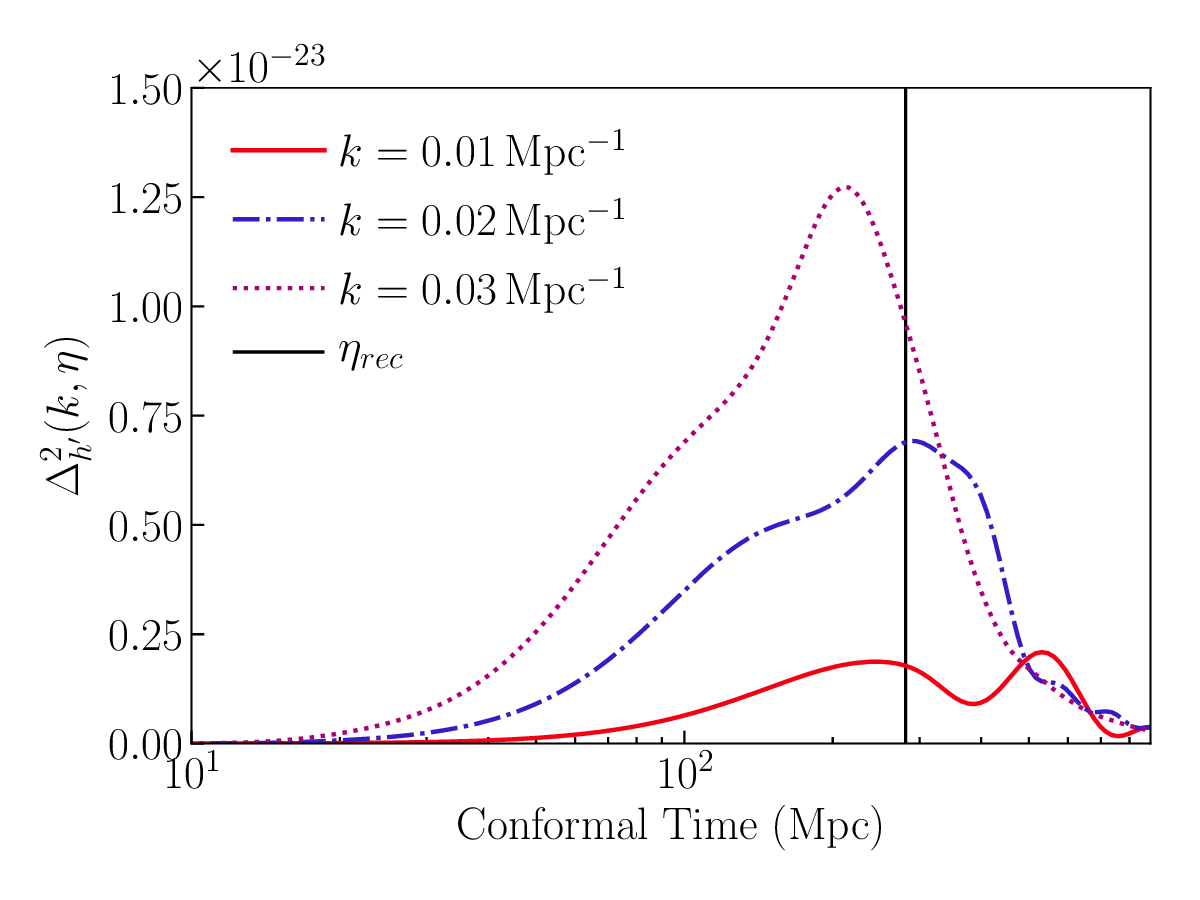}
    \end{subfigure}
}
\captionsetup{singlelinecheck = false,  justification=raggedright, labelsep=space}
    \caption{The evolution of $\Delta^2_h(k)$ (left panel) and $\Delta^2_{h'}(k)$ (right panel) for three relevant modes. The shape of $\Delta^2_{h'}(k)$ can be understood as a superposition of a term shaped like $\partial_\eta P_h(k)$ and a term proportional to the source.}
    \label{fig:evol}
\end{figure}

We show the results of this calculation on the left-hand side of Fig~\ref{fig:evol} as the dimensionless power spectrum $\Delta^2_h(k, \eta)\equiv k^3 P_h(k, \eta)/(2\pi^2)$. The dimensionless power spectrum quantifies the power per logarithmic interval, taking into account the Fourier-space volume. Note that while $P_h(k)$ is constant and non-zero on superhorizon limit, $k\to0$, $\Delta^2_h(k)$ is indeed suppressed on super-horizon scales.  We also note that the oscillatory behavior associated with gravitational waves freely propagating inside the horizon (which are periodic in $k\eta$) is largely absent in our signal. This is because at the scales of interest the oscillation due to the Green's function is dominated by the new gravitational wave amplitude currently being sourced.

The amplitude of $\Delta^2_h(k, \eta)$ is characteristically about 10 orders of magnitude smaller than the scalar perturbations $\Delta^2_\zeta \approx 10^{-9}$. We can estimate this amplitude from \eqref{eq:vsol} by writing
\ba
\Delta_h^2(k, \eta) 
\sim \frac{k^3}{2\pi^2}h^2 
\sim  
\frac{k^3}{2\pi^2}
\left[8 \pi G \rho_{\bc}(\eta_{\rm rec}) a(\eta_{\rm rec}){\cal G}(k, \eta_{\rm rec}, \eta_{\rm source}) \Delta \eta \beta^2\right]^2,
\ea
where $\beta = v_{\bc}/c$.
From the Friedmann equation, 
\be
8 \pi G \rho_{\bc} a(\eta)^2 =3 \Omega_{\bc} {\cal H}(\eta)^2,
\ee
where $3\Omega_{\bc} \sim 10^{-1}$ and in the radiation epoch (which is when the largest tensor modes are generated) ${\cal H} = 1/\eta \sim 10^{-2}$ Mpc. Finally, the Green's functions scales as $k^{-1}$ and since the source is not well localized temporally, we take $\Delta\eta = \eta \sim 10^2$ Mpc. With $k = 10^{-2}$ Mpc, we find $\Delta^2_h(k) \sim 10^{-19}$ which is within a couple orders of magnitude of the full result. It is clear from this estimate alone that the signal will not be detectable. We stress again, however, that this signal is physically observable in principle, as we shall show explicitly in the next section.

\section{B-mode polarization power spectrum}
The physical observable relevant to the IGW on these scales is the divergence-free (``B-mode'') part of the polarization of the cosmic microwave background. Exactly determining this quantity would require knowledge of the evolution $\Delta^2_h(k, \eta)$ for all time, which is computationally quite demanding. Instead, we adopt the tight-coupling approximation of \cite{pritchardcosmic2005} that assumes at early times Thomson scattering efficiently washes out all anisotropies, so that the temperature and polarization multipoles are generated entirely during the period of recombination. In this approximation the B-mode angular power spectrum is given as 
\be
C_{\ell}^{BB} =  \left(\frac{2}{7} \log\frac{10}{3}\right)^2 \pi\int \frac{dk}{k} \Delta^2_{h'}(k, \eta_r)
\[P_{\ell}^{BB}(k(\eta_0 - \eta_{\rm rec}))\]^2\(\Delta\eta_{\rm rec}\)^2 e^{-{(k\Delta \eta_{\rm rec})^2}}
\ee
where $\Delta \eta_{\rm rec}$ is the width of the visibility function (approximated as a Gaussian) in conformal time and $P_{\ell}^{BB}$ represents the projection factor from the Fourier basis to the B-mode multipoles \cite{TAM}, given by 
\be
P_\ell^{BB}(x) = j'_\ell(x)+\frac{2j_\ell(x)}{x},
\ee
with $x = k (\eta_0-\eta_{\rm rec})$ the look-back time from the present scaled by the wavenumber. Note that $P_{\ell}^{BB}$ is precisely the radial function relating spherical tensor basis $Y_{(JM)ab}^{TB}$ and the B-mode tensor Total-Angular-Momentum wave basis $\Psi^{k,TB}_{(JM)ab}$ in Ref.~\cite{TAM}. Ref.~\cite{pritchardcosmic2005} used a polynomial approximation for the projection factor, while we use the exact expression. 

We calculate $\Delta^2_{h'}(k,\eta)$ in the absence of the primordial linear gravitational waves ($c_1=c_2=0$ in \refeq{vsol}), for which case the derivative becomes
\be
h'(\vk, \eta) = -\frac{\mathcal{H} (\eta)}{a(\eta)}
\int_0^\eta
d\eta'\;
a(\eta') \cG(k,\eta,\eta')
s_p(\vk,\eta')
+
\frac{1}{a(\eta)}
\int_0^\eta
d\eta'\;
\partial_\eta\left[a(\eta') \cG(k,\eta,\eta')
s_p(\vk,\eta')\right]\,.
\ee
Another term involving ${\mathcal G}(k, \eta,\eta)$ from the Leibniz integral rule vanishes due to anti-symmetry of the Green's function [\refeq{anti-G}]. Analogous to our calculation of $P_h(k)$, we find
\ba
P_{h'}(k, \eta) 
=
&\mathcal{H}^2(\eta) P_h(k) 
+ \frac{8\pi}{a^2(\eta)} \bigg[ -2 \mathcal{H}(\eta) \int_0^\eta d\eta'\,a^3(\eta')\partial_\eta{\cal G}(k, \eta, \eta')\rho_{\bc}(\eta')\int_0^\eta d\eta''\,a^3(\eta''){\cal G}(k, \eta, \eta'') \rho_{\bc}(\eta'')
\vs
 &+\int_0^\eta d\eta'\,a^3(\eta')\partial_\eta{\cal G}(k, \eta, \eta')\rho_{\bc}(\eta')\int_0^\eta d\eta''\,a^3(\eta'')\partial_\eta{\cal G}(k, \eta, \eta'') \rho_{\bc}(\eta'') \bigg]
\vs
&\times
\int_{\vq}
\frac{\[{\eps}^{p,ij}(\widehat{\bm{z}})q_iq_j\]^2}{q^2|\vk-\vq|^2}
T_{\bc}(q,\eta')
T_{\bc}(|\vk -\vq|,\eta')
T_{\bc}(q,\eta'')T_{\bc}(|\vk -\vq|,\eta'')  
P_\zeta(q) P_\zeta(|\vk-\vq|)
\,.
\ea
\begin{figure}
    \includegraphics[width=\textwidth]{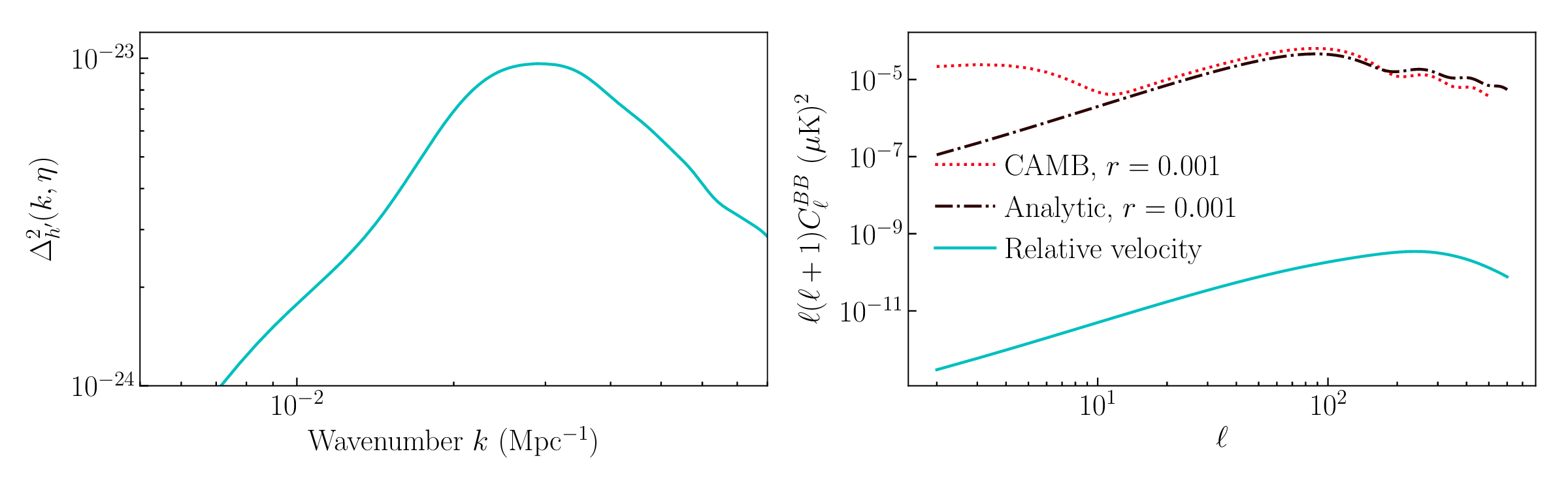}
    \captionsetup{singlelinecheck = false, justification=raggedright, labelsep=space}
    \caption{\textbf{Left:} The dimensionless power spectrum $\Delta^2_{h'}(k)$ at recombination. \textbf{Right:} The B-mode polarization from the primordial gravitational waves and the relative velocity. The primordial gravitational wave signal for a tensor-to-scalar ratio $r=0.001$ is calculated using CAMB and using the analytic method of \cite{pritchardcosmic2005}. Note that this analytic method does not account for the low-$\ell$ reionization bump. }
    \label{fig:c_ell}
\end{figure}
We show the time-evolution results of $\Delta^2_{h'}(k,\eta)$ in the right panel of FIG.~\ref{fig:evol}. Note that $\Delta^2_{h'}(k)$ cannot be visually interpreted as a derivative of the power spectrum $\Delta^2_h(k)$ because $\partial_\eta \Delta^2_h(k,\eta) \propto \langle 2 h h'\rangle$. Obviously, the maxima of $\Delta^2_h(k)$ do not correspond to zeros of $\Delta^2_{h'}(k)$ (except in the source free case). A more intuitive way of understanding the shape of $\Delta^2_{h'}$ is to return to the equation of motion:
\be
h'(k, \eta) = \frac{s_p(k, \eta) - k^2h(k, \eta) -  h''(k, \eta)}{2{\cal H} }\,.
\ee
In this form, it is clear that there are really two terms contributing to $h'$. The first is directly proportional to the source term, while the second represents the unforced damped harmonic oscillator response to the current value of $h(k,\eta)$. When we look at the evolution of $\Delta^2_{h'}(k)$ in the right hand side of  FIG.~\ref{fig:evol}, as we move to higher $k$ we are moving from a regime where the source term response dominates to one where the free-streaming response dominates. This is consistent with the larger values of $\Delta^2_h(k)$ for lower $k$ and the $k^3$ suppression of the relative velocity source. 

We show the dimensionless power spectrum $\Delta^2_{h'}(k)$ at recombination (left panel) and the B-mode power spectrum $C_\ell^{BB}$ (right panel) in FIG.~\ref{fig:c_ell}. As we have estimated from the previous section, unfortunately, at all scales the relative velocity IGW amplitude is several orders of magnitude too small to be observed even by next generation polarization experiments such as LiteBIRD \cite{Hazumi2020}, CMB stage-IV (e.g.~Simons Observatory \cite{Ade2019} and CMB-S4 \cite{abazajian2016cmbs4}), and PICO \cite{hanany2019pico}. Still, we can extract some physical meaning from this plot. We observe that $C_\ell^{BB}$ from the relative velocity IGW peaks at smaller scales than the primordial tensor modes. This can be attributed to two causes. First, the particle horizon exceeds the sound horizon by a factor of $\sqrt{3}$. Second, the primordial tensor modes are frozen outside the horizon, and begin to decay immediately at horizon entry. In contrast, the relative velocity IGW at first grows inside the horizon, and peaks significantly after horizon entry.

\section{Discussion}
We have calculated the dimensionless gravitational wave power spectrum $\Delta^2_{h'}(k)$ and CMB B-mode power spectrum from the gravitational waves induced by the relative velocity between baryons and cold dark matter (FIG.~\ref{fig:c_ell}). Although the dimensionful power spectrum $P_{h'}(k)$ is nonzero on superhorizon scales, causality is respected in $\Delta^2_{h'}(k)$ by the Fourier-space volume factor (which appears in all physical observables). We also calculated the time evolution of $\Delta^2_h(k)$ and $\Delta^2_{h'}(k)$ (the latter being responsible for generating the B-mode polarization of CMB) for several modes $k$ (FIG.~\ref{fig:evol}). The shape of $\Delta^2_{h'}(k)$ is explained as a sum of the freely propagating GW solution and the source term forcing. 

Although the effect we have studied is too small to be detectable in the foreseeable future, the physical interpretation is unambiguous. This is both because the source (the relative velocity) is gauge invariant, and because we have carried the calculation all the way to the relevant physical observable on the sky. This work does not address the general issues associated with the gauge freedom in nonlinear cosmological perturbation theory. Instead, it can be viewed as a ``minimal working example'' calculation of the physical consequences of an IGW effect. 

As a final exercise, we can consider the IGW from the relative velocity between protons and electrons due to the preferential scattering of photons off electrons. This effect has been previously considered for its ability to generate cosmological magnetic fields \cite{Harrison69,Harrison70,ichikicosmological2006}. Ref.~\cite{ichikicosmological2006} provides an estimate for the relative velocity $v\sim 10^{-2} k \left(\frac{1+z}{10^4}\right)^{-2}$. Substituting this into our own IGW estimate and noting that the ``reduced density'' for this problem is $\frac{\rho_e \rho_b}{\rho_e + \rho_b} \sim \rho_e$, we find a magnitude comparable to the baryon-CDM IGW and also certainly too small to observe. 

Calculating other IGW effects such as the density perturbation IGW will require greater care because the source may be gauge dependent. Arguments that the gauge dependence must disappear far inside the horizon \cite{Luca2020, inomatagauge2020, domenech2020approximate} have a close analog in linear perturbation theory. There, the expression for the density perturbation far inside the horizon coincides in many, but not all, gauges \cite{bardeengauge-invariant1980}. In that case, the issue has been resolved by realizing that we measure the density power spectrum using photons emitted from distant galaxies, so that the density perturbation should be measured through its effect on photon geodesics \cite{Yoo2009, Jeong2012, Challinor2011}. Similarly, to completely resolve this issue with the IGW it will be necessary to determine the frame relevant to the measurement. Only then can the correct gauge invariant variable relevant to the problem be defined.

\begin{acknowledgments}

The authors thank Misao Sasaki for the discussion on the existence of the zero-IGW gauge.
D.J.~and J.G.~were supported at Pennsylvania State University by NASA ATP program (80NSSC18K1103), and the Charles E. Kaufman Foundation of the Pittsburgh Foundation. J.H.~was supported by Basic Science Research Program through the National Research Foundation of Korea funded by the Ministry of Science (No.~2018R1A6A06024970 and 2019R1A2C1003031). H.N.~was supported by the National Research Foundation of Korea funded by the Korean Government (No.~2018R1A2B6002466)
\end{acknowledgments}

\appendix
\section{The zero-IGW gauge}\label{app:app}
Here, we show that as long as the IGW is gauge-dependent, we can always take a certain slicing where the IGW vanishes. For the pure scalar perturbation working as the source, the tracefree-transverse metric perturbation to the second-order, defined in (7), transform under the gauge transformation, see (32) and (36) in \cite{hwanggauge2017}, as
\bea
   & & \widehat h_{ij}
       = h_{ij} + {\cal P}_{ij}^{\;\;\;k\ell}
       \left( {2 \over a} \chi_{,(k} \xi^0_{\ell)}
       - \xi^0_{,k} \xi^0_{,\ell} \right),
\eea
where $g_{0i} \equiv - a \chi_{,i}$ and we have $\widehat \chi = \chi - a \xi^0$, see (24) in \cite{hwanggauge2017}. Let us assign the hat-frame the zero-IGW gauge (indicated by sub-$h$ with $h_{ij h} \equiv 0$) and the non-hat-frame the zero-shear (Poisson) gauge (indicated by sub-$\chi$ with $\chi_\chi \equiv 0$). Thus, we have $\chi_h = - a \xi^0_{\chi \rightarrow h}$ and
\bea
   & & h_{ij\chi} = {\cal P}_{ij}^{\;\;\;k\ell}
       \left(
       \xi^0_{{\chi \rightarrow h},k} \xi^0_{{\chi \rightarrow h},\ell} \right)
       = {1 \over a^2} {\cal P}_{ij}^{\;\;\;k\ell}
       \left( \chi_{h,k} \chi_{h,\ell} \right).
\label{eq:zeroIGWeq}
\eea
Here's an example procedure of how one can find the function $\chi$. 
Let us take a scalar Fourier transformation:
\be
\chi(\vx) = \int\frac{d^3q}{(2\pi)^3} \chi(\vq) e^{i\vq\cdot\vx}\,,
\ee
and try to find a relationship between $\chi(\vq)$ and $h_{ij}(\vk)$ to satisfy the
zero-IGW condition:
\be
h_{ij}(\vk) = 
\frac1{a^2}{{\cal P}_{ij}}^{k\ell}(\vkhat)[\chi_{,k}\chi_{,\ell}](\vk)\,.
\ee
Equivalently,
\ba
h_p(\vk)
=&
-\frac{1}{2a^2}
\varepsilon^{p,ij}(\vkhat)
{{\cal P}_{ij}}^{k\ell}(\vkhat)
\int\frac{d^3q}{(2\pi)^3}
q_k (k-q)_\ell
\chi(\vq)\chi(\vk-\vq)
\vs
=&
\frac{1}{2a^2}
\int\frac{d^3q}{(2\pi)^3}
\left[
\varepsilon^{p,k\ell}(\vkhat) q_k q_\ell
\right]
\chi(\vq)\chi(\vk-\vq)
\vs
=&
\frac{1}{2a^2}
\int\frac{d^3q}{(2\pi)^3}
q^2 \sin^2\theta
\left(
\begin{array}{cc}
    \cos(2\varphi) & ,~p=+\\
    \sin(2\varphi) & ,~p=\times
\end{array}
\right)
\chi(q,\theta,\varphi)
\chi(\sqrt{k^2-q^2-2kq\cos\theta},{\rm tan}^{-1}\(\frac{-q\sin\theta}{k-q\cos\theta}\),\varphi)\,.
\label{eq:zeroIGWsoln}
\ea
To obtain the second equality, we use that 
${\varepsilon}^{p,k\ell}(\khat)k_\ell=0$, and the third equality takes 
$k$ to $+\qhat_z$ direction without loss of gnerality.
Although we do not pursue a closed form solution here, we anticipate that 
one can find the solution $\chi(\vq)$ satisfying (A5) for a general $h_p(\vk)$,
as long as $h_p(\vk)$ only depends on $k$, a condition imposed by 
that the right-hand-side of \refeq{zeroIGWsoln} only depends on $k$.
That IGW only depends on $k$ is a characteristic of the IGW induced by scalar fields \cite{hwanggauge2017}, so we surmise that it must be possible to find a $\chi(\vq)$ with a full three-dimensional dependence for a given one-dimensional function $h_p(\vk)\equiv h_p(k)$.
Based on the $\varphi$-dependence of $\varepsilon^{p,k\ell}(\khat)q_kq_\ell$, 
we deduce that the scalar function $\chi(\vq)$ must have non-zero 
$Y_{\ell\pm1}(\qhat)$ modes.
Ref.~\cite{domenech2020approximate} has found a solution $\chi(\vq,\vk)$ that satisfy \refeq{zeroIGWeq} for a single Fourier mode.

The choice of $\chi_h$ (the zero-IGW gauge) satisfying the above relation makes the IGW $h_{ij}$ in that gauge vanish. This, however, does not imply that the IGW can be ``gauged away" as mentioned in \cite{domenech2020approximate}; $h_{ij\chi}$ is the same as $h_{ij}$ in the zero-shear gauge and indicates a gauge-invariant combination, thus cannot be gauged away. It merely reflects that the IGW is gauge-dependent, and therefore we can find the gauge condition where the IGW vanishes.

For example, the density perturbation is gauge-dependent in linear perturbation, and this is why we can choose the uniform-density gauge (setting $\delta \rho \equiv 0$) as a legitimate gauge condition. This is possible even in the presence of some density perturbation. As this gauge condition completely fixes the gauge degree of freedom, all the other variables in this gauge are in fact gauge invariant. There are infinitely many (as the slicing hypersurface can be continuously deformed) gauge conditions possible, which leave the remaining variables all gauge invariant. ``Which is the right gauge choice to describe the density perturbation?" should be based on the observational strategy as performed in Refs.~\cite{Yoo2009,Challinor2011, Jeong2012} for the matter power spectrum. The same must apply to the gauge-dependent IGW. Our study is free from this ambiguity because the baryon-cold dark matter velocity difference, the source of IGW in this work, is intrinsically gauge-invariant.

\bibliographystyle{apsrev4-2}
\bibliography{vbc}

\end{document}